\documentstyle[12pt]{article}

\newcommand{\beq}{\begin{equation}}
\newcommand{\eeq}{\end{equation}}
\def\bea{\begin{eqnarray}}
\def\eea{\end{eqnarray}}

\def\sss{\scriptscriptstyle}

\def\bd{B_d^0}
\def\bdbar{{\overline{B_d^0}}}
\def\bs{B_s^0}
\def\bsbar{{\overline{B_s^0}}}
\def\barp{{\raise.35ex\hbox{${\sss (}$}}---{\raise.35ex\hbox{${\sss )}$}}}
\def\bdbarp{\hbox{$B_d$\kern-1.4em\raise1.4ex\hbox{\barp}}}
\def\bsbarp{\hbox{$B_s$\kern-1.4em\raise1.4ex\hbox{\barp}}}
\def\ks{K_{\sss S}}

\def\kbar{{\overline{K^0}}}

\def\roughly#1{\mathrel{\raise.3ex\hbox{$#1$\kern-.75em\lower1ex\hbox{$\sim$}}}}

\def\Abar{\bar A}

\def\epjc#1#2#3{{\it Eur.\ Phys.\ J.}\ {\bf C#1}, #3 (19#2)} 

\def\npb#1#2#3{{\it Nucl.\ Phys.}\ {\bf B#1}, #3 (19#2)}
\def\plb#1#2#3{{\it Phys.\ Lett.}\ {\bf #1B}, #3 (19#2)}
\def\prd#1#2#3{{\it Phys.\ Rev.}\ {\bf D#1}, #3 (19#2)}
\def\prl#1#2#3{{\it Phys.\ Rev.\ Lett.}\ {\bf #1}, #3 (19#2)}
\def\zpc#1#2#3{{\it Zeit.\ Phys.}\ {\bf C#1}, #3 (19#2)} 
\def\ijmp#1#2#3{{\it Int.\ J.\ Mod.\ Phys.}\ {\bf A#1}, #3 (19#2)}

\begin{document}

\begin{flushright}
UdeM-GPP-TH-00-66 \\
IMSc-00/02/05 \\
\end{flushright}

\begin{center}
\bigskip

{\Large \bf Is it possible to Measure the Weak Phase}\\ 
{\Large \bf of a Penguin Diagram?}\footnote{Talk given by Rahul
Sinha. To appear in Proceedings of the Third International Conference
on B Physics and CP Violation, Taipei, December 3-7, 1999, H.-Y. Cheng
and W.-S. Hou, eds.  (World Scientific, 2000).}  \\
\bigskip

David London $^{a,}$\footnote{london@lps.umontreal.ca},~~ 
Nita Sinha $^{b,}$\footnote{nita@imsc.ernet.in}~~
and Rahul Sinha $^{b,}$\footnote{sinha@imsc.ernet.in}
\end{center}


\begin{flushleft}
~~~~~~~~~~~$a$: {\it Laboratoire Ren\'e J.-A. L\'evesque, 
Universit\'e de Montr\'eal,}\\
~~~~~~~~~~~~~~~{\it C.P. 6128, succ. centre-ville, Montr\'eal, QC,
Canada H3C 3J7}\\
~~~~~~~~~~~$b$: {\it Institute of Mathematical Sciences, Taramani,
 Chennai 600113, India}
\end{flushleft}

\begin{center}
 
\bigskip (\today)

\bigskip 

{\bf Abstract}

\end{center}

\begin{quote}

 The $b\to d$ penguin amplitude receives contributions from
internal $u$, $c$ and $t$-quarks. We show that it is impossible to
measure the weak phase of any of these penguin contributions without
theoretical input. However, a single assumption involving the hadronic
parameters makes it possible to obtain the weak phase and test for the
presence of new physics in the $b\to d$ flavour-changing neutral
current.
\end{quote}
\newpage

\section{Introduction}
One of the most compelling features of CP violation in the $B$ system
is that all three interior angles of the unitarity triangle,
$\phi_1(\equiv\beta)$, $\phi_2(\equiv\alpha)$ and
$\phi_3(\equiv\gamma)$\cite{pdg}, can be measured cleanly, i.e.\
without theoretical hadronic uncertainties. The $B$ system is thereby
expected to provide a test of CP violation in the standard model
(SM). Any inconsistency with the predictions of the SM will reveal the
much sought after signal of new physics (NP).

NP can affect CP violation in one of two possible ways: through
contributions to $B$ decays or to $\bd$-$\bdbar$ mixing. Most decay
modes of the $B$-meson are dominated by $W$-mediated tree-level
diagrams and will not be much affected by NP, since in most models of
NP there are no contributions that can compete with the SM. Thus, with
the exception of penguin-dominated decay modes, NP cannot
significantly affect the decays. However, new contributions to
$\bd$-$\bdbar$ mixing can affect the CP
asymmetries\cite{NPBmixing}. Such NP contributions will affect the
extraction of $V_{td}$ and $V_{ts}$, as well as possible measurements
of $\phi_1,\phi_2$ and $\phi_3$. Thus, NP enters principally through
contributions to $\bd$-$\bdbar$ mixing\cite{Bnewphysics}.

The angles $\phi_1$,$\phi_2$ and $\phi_3$ are to be measured
principally through the modes $\bd(t) \to \Psi\ks$, $\bd(t) \to \pi
\pi$ (or $\rho\pi$)\cite{isospin}, and $B^\pm \to D K^\pm$ (or $ D^*
K^{*\pm}$)\cite{BtoDK}, respectively. NP in $\bd$-$\bdbar$ mixing will
then affect the measurements of $\phi_1$ and $\phi_2$, but in opposite
directions\cite{NirSilv}. That is, in the presence of a new-physics
phase $\phi_{\sss NP}$, the CP angles are changed as follows: $\phi_1
\to \phi_1 - \phi_{\sss NP}$ and $\phi_2 \to \phi_2 + \phi_{\sss NP}$.
Hence the sum $\phi_1 + \phi_2 + \phi_3$ is {\it insensitive} to the
NP. However, if $\phi_3$ is measured in the decay $\bs(t) \to D_s^\pm
K^\mp$ \cite{BstoDsK}, then $\phi_1 + \phi_2 + \phi_3 \ne \pi$ can be
found if there is NP in $\bs$-$\bsbar$ mixing.

The most well known method for detecting NP is to compare the unitary
triangle as constructed from measurements of the angles with that
constructed from independent measurements of the sides. Any
inconsistency will be evidence for new physics. However, since at
present the allowed region of the unitarity triangle is rather large,
the triangle as constructed from the angles could still lie within the
allowed region even if NP is present.  Furthermore, even if the
$\phi_1$-$\phi_2$-$\phi_3$ triangle lies outside the allowed region,
one might still be skeptical about the presence of NP: perhaps the
theoretical uncertainties which go into the constraints on the
unitarity triangle have been underestimated.

Clearly we would like cleaner, more direct tests of the SM in order to
probe for the presence of NP. More promising tests for NP are possible
by comparing two distinct decay modes which, in the SM, probe the same
CP angle. One can compare the rate asymmetries in $B^\pm \to D K^\pm$
and $\bs(t) \to D_s^\pm K^\mp$, both of which measure $\phi_3$. A
discrepancy between the extracted values would point to NP in
$\bs$-$\bsbar$ mixing. Similarly, a discrepancy in $\phi_1$, as
measured via $\bd(t) \to \Psi\ks$ and $\bd(t) \to \phi\ks$, implies
new physics in the $b\to s$ penguin\cite{NPpenguins}. One can also
measure the CP asymmetry in the decay $\bs(t)\to\Psi\phi$, which
vanishes to a good approximation in the SM. Such an asymmetry would
indicate the presence of new physics in $\bs$-$\bsbar$ mixing. Note
that all such tests probe NP in the $b\to s$ flavour-changing neutral
current (FCNC).

One may then ask the question: are there are any direct tests of NP in
the $b\to d$ FCNC? For example, consider pure $b\to d$ penguin decays
such as $\bd \to K^0\kbar$ or $\bs\to\phi\ks$, with the asumption that
$t$-quark contribution dominates among up-type quarks in the loop. In
such a case the SM would predict that (i) the CP asymmetry in $\bd(t)
\to K^0\kbar$ vanishes, and (ii) the CP asymmetry in $\bs(t)
\to\phi\ks$ measures $\sin 2\phi_1$ \cite{penguins}. Any discrepancy
between measurements of these CP asymmetries and their predictions
would thus imply that there is NP in either $\bd$-$\bdbar$ mixing or
the $b\to d$ penguin, i.e.\ in the $b \to d$ FCNC. However, it is well
known that $b\to d$ penguins are {\it not} dominated by the internal
$t$-quark. The contributions of the $u$- and $c$-quarks can be as
large as 20--50\% of that of the $t$-quark\cite{ucquark}. As a
consequence, one cannot probe NP in $b\to d$ FCNC using such modes,
and, unfortunately, the answer to the question asked is \textbf{\it
no}\cite{LSS}.

\section{The CKM Ambiguity}

The full $b\to d$ penguin amplitude is a sum of contributions from the
three internal up-type quarks in the loop: 
\beq 
P = P_u \, V_{ub}^*
V_{ud} + P_c \, V_{cb}^* V_{cd} + P_t \, V_{tb}^* V_{td} ~,
\label{bdpenguin}
\eeq
with $V_{ub} \sim e^{-i \phi_3}$ and $V_{td} \sim e^{-i \phi_1}$.
Using the unitarity relation, $V_{ud} V_{ub}^* + V_{cd} V_{cb}^* +
V_{td} V_{tb}^* = 0$, the $u$-quark piece can be eliminated in
Eq.~(\ref{bdpenguin}), allowing us to write
\beq P = {\cal P}_{cu} \, e^{i\delta_{cu}} + {\cal P}_{tu} \,
e^{i\delta_{tu}} e^{-i \phi_1 }~,
\label{peng1}
\eeq
where $\delta_{cu}$ and $\delta_{tu}$ are strong phases. Now imagine
that there were a method in which a series of measurements allowed us
to cleanly extract $\phi_1$ using the above expression. In this case,
we would be able to express $-\phi_1$ as a function of the
observables.

On the other hand, we can instead use the unitarity relation to
eliminate the $t$-quark contribution in Eq.~(\ref{bdpenguin}),
yielding
\beq
P = {\cal P}_{ct} \, e^{i\delta_{ct}} + {\cal P}_{ut} \, e^{i\delta_{ut}} 
e^{i \phi_3} ~.
\label{peng2}
\eeq
Comparing Eqs.~(\ref{peng1}) and (\ref{peng2}), we see that they have
the same form. Thus, the same method used to extract $-\phi_1$ from
Eq.~(\ref{peng1}) can be used on Eq.~(\ref{peng2}) to obtain
$\phi_3$. That is, we would be able to write $\phi_3$ as {\it the same
function} of the observables as was used for $-\phi_1$ above! But this
implies that $-\phi_1 = \phi_3$, which clearly does not hold in
general.

Due to the ambiguity in the parametrization of the $b\to d$ penguin
--- which we refer to as the {\it CKM ambiguity} --- we conclude that
one cannot cleanly extract the weak phase of any penguin contribution.
Indeed, it is {\it impossible} to cleanly test for the presence of new
physics in the $b\to d$ FCNC.  Nevertheless, it is instructive to
examine in detail a few candidate methods, to see exactly how they
fail.


The measurement of the time-dependent rate for the decay $\bd(t) \to
K^0\kbar$ can at best allow one to extract the magnitudes and relative
phase of $e^{i\phi_1} A$ and $e^{-i\phi_1} \Abar$, where $A$ is the
amplitude for $\bd \to K^0\kbar$. With an independent measurement of
$\phi_1$, there are a total of 4 measurements. Using the form of the
$b\to d$ penguin given in Eq.~\ref{peng1}, we have
%
$
  e^{i\phi_1} A = e^{i\phi_1} 
(
{\cal P}_{cu} \, e^{i\delta_{cu}} 
+ {\cal P}_{tu} \, e^{i\delta_{tu}} e^{-i \phi_1'} 
)
$, where we have written the phase $\phi_1'$ to allow for the
possibility of new physics.
%
There are thus 5 theoretical (hadronic) parameters: ${\cal P}_{cu}$,
${\cal P}_{tu}$, $\delta_{cu}-\delta_{tu}$, $\phi_1$, and
$\theta_{\sss NP} \equiv \phi_1'-\phi_1$. We see that there are not
enough measurements to determine all the theoretical parameters. In
fact, there is just one more theoretical unknown than there are
measurements.  A similar examination\cite{LSS} of the $B\to \pi\pi$
isospin analysis, Dalitz-plot analysis of $B \to 3\pi$, angular
analysis of $B^0 \to V V$ (where $V$ is a vector meson),
and a combined isospin $+$ angular analysis of $B\to \rho\rho$ leads
to the same conclusion that there is one more unknown than there are
measurements.

We thus conclude that, due to the CKM ambiguity, if one wishes to test
for the presence of NP in the $b\to d$ FCNC by comparing the weak
phase of the $t$-quark contribution to the $b\to d$ penguin with that
of $\bd$-$\bdbar$ mixing, it is necessary to make a single
assumption\cite{KLY2} about the hadronic parameters.



%

\section*{Acknowledgments}
R.S. would like to thank the organizers of this conference,
Prof. H.Y. Cheng and Prof. A.I. Sanda, for financial assistance to
attend the conference. 
The work of D.L. was financially supported by NSERC of Canada. 

\end{document}